# Spin-Dependent Charge-State Conversion in NV Ensembles Mediated by Electron Tunneling

Neil B. Manson[1,*], Morgan Hedges[1], Michael S. J. Barson[2], Carlos A. Meriles[3,4], Ronald Ulbricht[5], and Marcus W. Doherty[1]

[1] Department of Quantum Science and Technology, Research School of Physics, Australian National University, Canberra, A.C.T. 2601, Australia.

[2] School of Physics and Astronomy, Monash University, Clayton, Vic 3800, Australia.

[3] Department of Physics, CUNY - City College of New York, New York, NY 10031, USA.

[4] CUNY - Graduate Center, New York, NY 10016, USA.

[5] Max Plank Institute for Polymer Research, Ackermannweg, 10, 55128 Mainz, Germany.

[*] E-mail: neil.manson@anu.edu.au

The nitrogen–vacancy (NV) center in diamond enables optical initialization and readout of its electronic spin, forming the basis of a wide range of quantum sensing and metrology applications. A central challenge in such measurements is the coexistence of two charge states, $NV^-$ and $NV^0$: While detection protocols rely on the spin-dependent properties of $NV^-$, fluorescence from $NV^0$ does not carry useful contrast and is typically removed as background, reducing the available signal. Here, we show that the origin of $NV^0$ emission depends strongly on the excitation wavelength in nitrogen-containing diamond. Using ensembles of NV centers with varying nitrogen concentrations, we compare excitation at the $NV^0$ zero-phonon line (ZPL) at 575 nm with the commonly used 532 nm. We find that excitation at 575 nm generates $NV^0$ predominantly through spin-selective tunneling from the excited state of $NV^-$ to nearby nitrogen donors, such that the $NV^0$ emission follows the spin polarization of $NV^-$. As a result, the $NV^0$ fluorescence contributes to the measurable spin contrast, allowing the full fluorescence signal to be used for detection. This result opens opportunities for improved sensitivity in NV-based sensing applications.

## I. INTRODUCTION

The nitrogen–vacancy (NV) center in diamond has attracted considerable attention because its electronic spin can be optically initialized and read out even at room temperature[1,2]. These properties have enabled a wide range of sensing applications[3], including nanoscale magnetometry[4-6] and electrometry[7,8], as well as thermometry[9-11] and pressure sensing[12]. In many of these applications the use of ensembles of NV centers is advantageous because larger numbers of defects can provide stronger fluorescence signals and improved measurement sensitivity[13-15]. A persistent complication in ensemble measurements, however, is the coexistence of two charge states of the defect: the negatively charged $NV^-$ center and the neutral $NV^0$ center[16]. Most sensing protocols rely on the optical and spin properties of $NV^-$, whereas $NV^0$ contributes fluorescence that does not provide useful spin-dependent information. As a result, $NV^0$ emission is often treated as an unwanted background and removed from the detected signal, reducing the available fluorescence and degrading the signal-to-noise ratio.

In nitrogen-containing diamond, charge-state dynamics are strongly influenced by the presence of substitutional nitrogen donors[17-19]. Electron tunneling between NV centers and nearby nitrogen impurities can convert $NV^-$ to $NV^0$ and vice versa, with rates that depend sensitively on the separation between defects[20]. Further, optical excitation can also modify the balance between charge states through processes that involve either tunneling or photoionization of donors.

In this work we show that the dominant mechanism producing $NV^0$ depends strongly on the excitation wavelength. Using ensembles of NV centers in diamonds with varying nitrogen concentrations, we compare excitation at 575 nm and the commonly used 532 nm wavelength. We find that 575 nm excitation generates $NV^0$ primarily through spin-selective tunneling from the excited state of $NV^-$ to nearby nitrogen donors, producing $NV^0$ emission that follows the spin polarization of $NV^-$. This is not the case for 532 nm excitation where the $NV^0$ fluorescence is insensitive to the $NV^-$ spin polarization. This distinction has important practical implications as spin-selective measurements can potentially benefit from contributions to the emission spectrum arising from both NV charge states, hence opening opportunities for improved sensitivity.

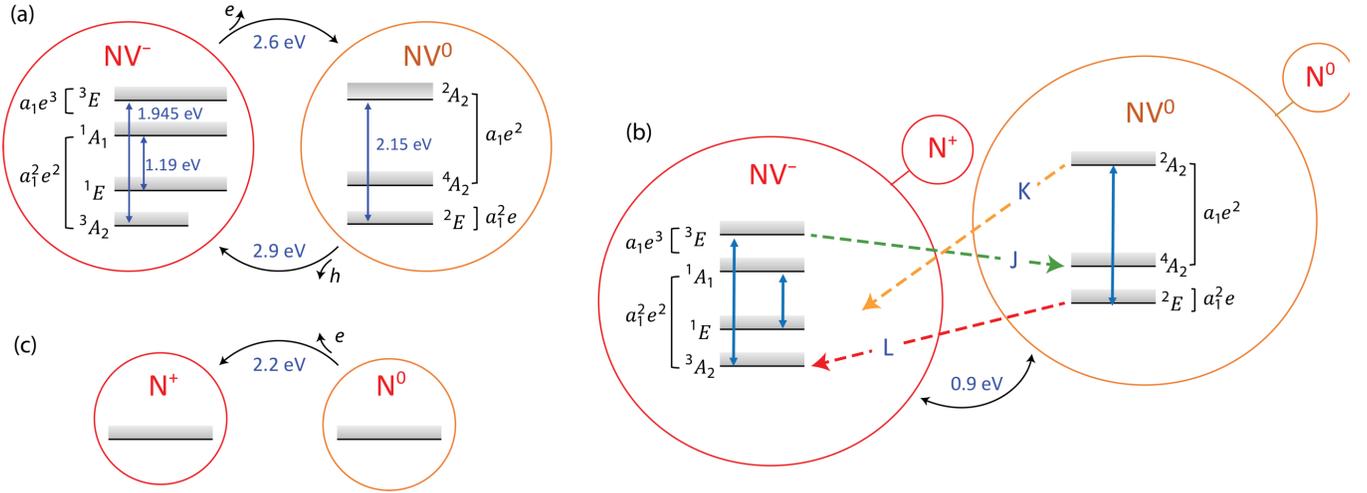

**Figure 1. Energy-level structure and charge-transfer pathways of NV centers in the absence and presence of nearby nitrogen donors.** (a) Energy levels of isolated NV⁻ and NV⁰ centers in nitrogen-poor diamond, showing optical transitions and the energies required for one-photon charge-state conversion. Charge transfer between NV⁻ and NV⁰ in this limit proceeds via band-mediated processes involving injection of an electron into the conduction band or the creation of a hole in the valence band. (b) Coupled NV–N system in nitrogen-containing diamond. When a substitutional nitrogen donor is located near an NV center, the energy separation between the configurations NV⁻/N⁺ and NV⁰/N⁰ is reduced (≈0.9 eV), enabling direct electron tunneling between the defects without involvement of band carriers. The coupling strength, and hence the tunneling rate, depends strongly on the NV–N separation. Representative tunneling processes are indicated: J, transfer from the excited state of NV⁻ to NV⁰; K, transfer from the excited state of NV⁰ to NV⁻; and L, spontaneous relaxation from NV⁰ to NV⁻. These processes govern the charge-state dynamics discussed in the text. (c) Ionization of substitutional nitrogen, illustrating the conversion of N⁰ to N⁻ with an energy of approximately 1.7-2.2 eV and the release of an electron to the conduction band.

## II. RESULTS

Before examining how optical excitation creates NV⁰, it is useful to recall the charge-transfer processes expected for isolated NV centers. Figure 1a summarizes the relevant electronic levels for NV⁻ and NV⁰ and the transitions — taking the form of one-photon[21] (or two-photon[22]) processes — that can change the charge state through excitation involving the conduction or valence bands. For isolated centers, conversion between NV⁻ and NV⁰ typically involves promoting an electron to the conduction band or creating a hole in the valence band. In other words, charge-state conversion in the single-center picture proceeds through band-mediated processes that involve free carriers[23-27] (Fig. 1a). This description is widely used to interpret charge dynamics in nitrogen-poor diamond but does not necessarily apply in samples containing significant concentrations of substitutional nitrogen donors.

Indeed, in such samples the situation is different: As illustrated in Fig. 1b, when a nitrogen donor lies close to an NV center the two defects can be seen as a coupled system in which charge transfer occurs through direct tunneling rather than through the bands[28]. In this case the relevant configurations are NV⁻/N⁺ and NV⁰/N⁰, whose energy separation is significantly reduced compared with the isolated-center case. Electron transfer between the NV and the donor can therefore occur without the involvement of conduction-band electrons or valence-band holes[20].

Three tunneling processes are particularly important for interpreting the experiments that follow. First, even in the absence of illumination, an electron can tunnel from a neutral nitrogen donor to NV⁰, producing NV⁻ and N⁺. This transition, labeled L in Fig. 1c, determines the equilibrium charge-state balance in donor-rich crystals[29]. Second, when NV⁻ is optically excited, an electron can tunnel from the excited NV⁻ center to a nearby donor, converting NV⁻ to NV⁰; this process is labeled J[30,31]. Finally, when NV⁰ is excited, tunneling in the reverse direction can occur, restoring NV⁻; this process is labeled K. The rates of these processes depend strongly on the separation between the NV and the donor, so ensemble measurements naturally reflect a distribution of tunneling rates[32]. For future reference, Fig. 1c also illustrates the ionization of substitutional nitrogen, which will become relevant when considering excitation at higher photon energies.

The influence of the donor concentration on the charge-state balance is illustrated in Fig. 2a. The figure shows emission spectra measured at 77 K under weak 532 nm illumination, with the intensity chosen so that the excitation does not significantly perturb the underlying

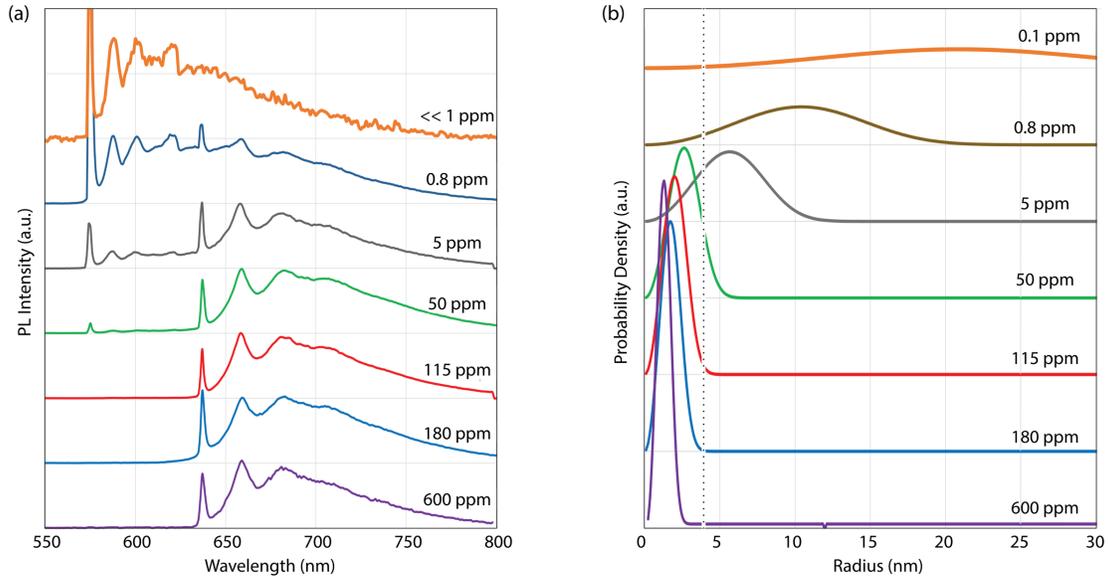

**Figure 2. L-tunnelling and its dependence on nitrogen concentration.** (a) Emission spectra at 77 K under low-intensity 532 nm excitation. Emission between 575 nm and 637 nm is due to NV⁰, whereas beyond 637 nm it is predominantly NV⁻. The relative NV⁻/NV⁰ populations are largely set in the dark by tunneling process L. (b) Calculated distribution of nearest-neighbor nitrogen distances for the corresponding donor concentrations. The dashed line at 4 nm marks a separation at which the NV⁻/NV⁰ balance changes markedly. All experiments at 77 K.

populations. The low temperature ensures clear spectral separation of the charge states, so that emission between ~575 nm and ~637 nm can be attributed to NV⁰, while emission on the long-wavelength side of the NV⁻ zero-phonon line (ZPL) at 637 nm is dominated by NV⁻. At low nitrogen concentrations, the paucity of donors implies that the electron reservoir is limited and the Fermi level is not well defined, so that the NV charge state need not correspond to a well-defined equilibrium[33]. In this regime, the charge state can become history-dependent, a property that underlies the use of NV centers for charge-based information storage in diamond[34,35]. As the nitrogen concentration increases, the relative NV⁻ emission rises correspondingly. This behavior reflects the growing probability that an NV center has a nearby nitrogen donor capable of supplying an electron through tunneling process L. The distribution of nearest-neighbor donor distances shown in Fig. 2b emphasizes this point: when donors occur within a few nanometers of the NV center, the tunneling rate becomes

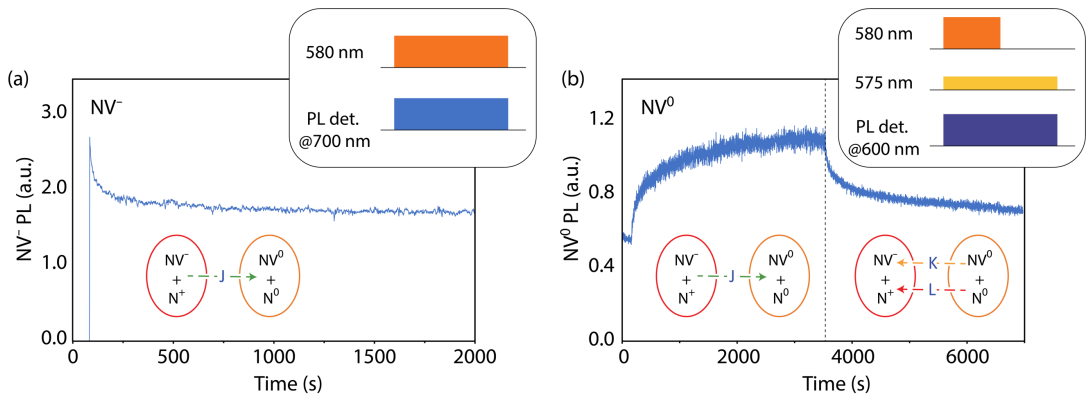

**Figure 3. Time-dependent charge-state dynamics under wavelength-selective excitation.** (a) NV⁻ emission (detected at 700 nm) under 580 nm excitation for the 115-ppm-nitrogen diamond. The signal exhibits a rapid initial drop followed by a slower decay, reflecting loss of NV⁻ through tunneling transition J (electron transfer from the excited state of NV⁻ to a nearby nitrogen donor), which converts NV⁻ to NV⁰. (b) NV⁰ emission (detected at 600 nm) under the same excitation conditions, measured using a weak 575 nm probe. The NV⁰ signal increases correspondingly, confirming charge transfer from NV⁻ to NV⁰ via transition J. When the 580 nm excitation is switched off (vertical dashed line), the system relaxes back to the initial state through the dark tunneling process L, which restores NV⁻.

sufficiently large that NV⁻ is strongly favored[20].

Having established the equilibrium situation, we now consider how optical excitation modifies the charge state through the tunneling processes described above. Figure 3 provides a direct demonstration using time-dependent measurements for the diamond with 115 ppm nitrogen content. When the sample is excited at 580 nm and the NV⁻ emission is monitored near 700 nm, the fluorescence intensity decreases with time. This decrease indicates that NV⁻ centers are gradually converted to NV⁰ through tunneling process J, in which an electron transfers from the excited state of NV⁻ to a nearby nitrogen donor. The decay is not a single exponential because different NV–N separations produce different tunneling rates[32]. Independent measurements detecting NV⁰ emission near 600 nm under weak 575 nm excitation confirm that the NV⁰ population increases as the NV⁻ signal decreases. When the 580 nm light is removed, the system slowly returns to its initial state through the dark tunneling process L, which again favors NV⁻.

These observations suggest that NV⁰ produced under excitation near 575 nm originates primarily from tunneling from the excited state of NV⁻. A decisive test of this interpretation is provided by the magnetic-field dependence of the emission. Optical pumping of NV⁻ produces spin polarization, which enhances the fluorescence intensity. Applying a sufficiently strong (>10-20 mT) off-axis magnetic field mixes the spin states and decreases this polarization, thereby reducing the fluorescence[1,36-40]. If NV⁰ is generated through tunneling from the excited state of NV⁻, its emission should follow the same magnetic-field dependence as the NV⁻ emission.

Figure 4a shows that this is in fact the case. Under 575 nm excitation the NV⁰ emission varies with magnetic field in exactly the same way as the NV⁻ emission. Indeed, when the magnetic field is applied both signals decrease together. This correlation indicates that the NV⁰ emission originates from the population of the NV⁻ excited state and therefore inherits the same spin-dependent modulation. In other words, NV⁰ produced under 575 nm excitation is generated predominantly through the tunneling process J.

The situation changes markedly for excitation at 532 nm, which is the wavelength most commonly used in NV experiments (Fig. 4b). Although this wavelength also excites NV⁻ and produces spin polarization, the photon energy is sufficiently large to ionize substitutional nitrogen donors[41-46] (see also Fig. 1c). As a result, NV⁰ is not produced predominantly through tunneling from the NV⁻ excited state. Instead, 532 nm excitation modifies the charge balance of the defect environment[47] and shifts the equilibrium toward NV⁰. Because this process does not originate from the NV⁻ excited state, the resulting NV⁰ fluorescence is expected to be largely insensitive to spin polarization and therefore independent of magnetic field, as observed in the spectra of Fig. 4b.

Although the samples we investigated span a wide range of nitrogen concentrations — from a few ppm to several hundred ppm — we consistently find they all display a qualitatively similar trend both at 575 and 532 nm excitation. The apparent magnitude of the effect at 532 nm, however, varies between samples, reflecting differences in the balance between the forward tunneling transition J and the back-transfer process L. In samples with high nitrogen concentrations the tunneling rates are large because nitrogen donors are typically located close to the NV centers. As a consequence, any $NV^0$ created through the J process rapidly converts back to $NV^-$ through the spontaneous transition L. This rapid back-transfer suppresses the steady-state $NV^0$ population in continuous-wave measurements even though the tunneling itself is efficient. Conversely, in samples with more moderate nitrogen concentrations the larger NV–N separations slow the back-transfer process, allowing $NV^0$ populations to accumulate and leading to stronger observed field-dependent $NV^0$ emission.

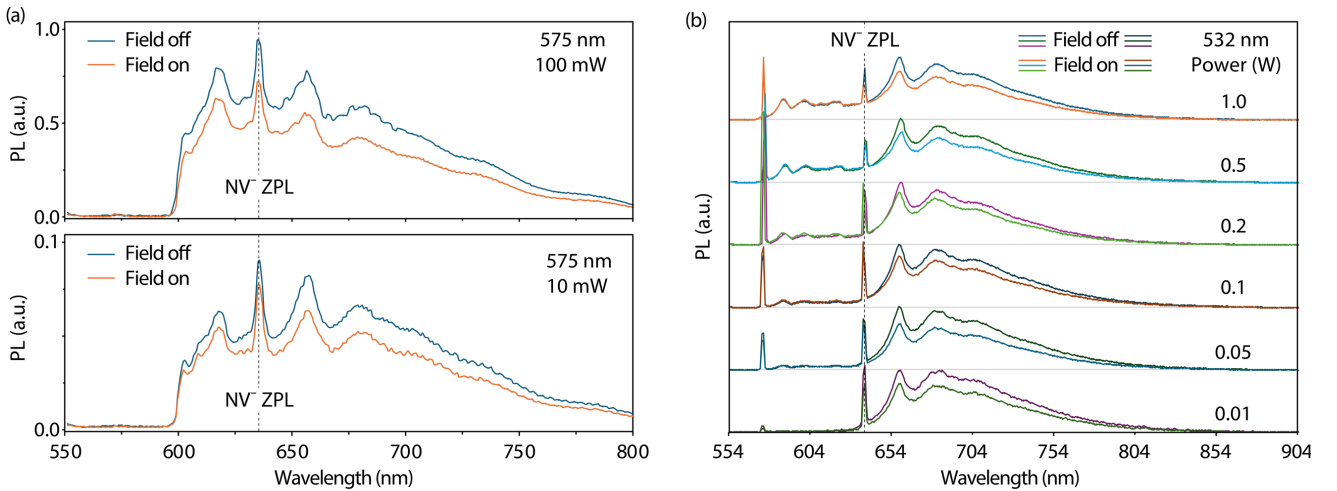

**Figure 4. Excitation-wavelength dependence of NV⁰ emission and its relation to spin polarization.** (a) Emission spectra under 575 nm excitation recorded with and without an applied magnetic field at two excitation powers. The NV⁰ emission — dominant below the NV⁻ ZPL at 637 nm — exhibits the same magnetic-field dependence as NV⁻, indicating that it follows its spin polarization. (b) Emission spectra under increasing 532 nm excitation intensity. In contrast to (a), the NV⁰ emission shows little or no dependence on the applied magnetic field. All experiments at 77 K using the 115 ppm diamond.

### III. CONCLUSION

In this work, we have shown that the mechanism responsible for the generation of $NV^0$ in nitrogen-containing diamond depends critically on the excitation wavelength. By comparing excitation at 575 nm and 532 nm across samples with varying nitrogen concentrations, we identify two distinct regimes of charge-state dynamics. Excitation near 575 nm produces $NV^0$ predominantly through tunneling from the optically excited state of $NV^-$ to nearby nitrogen donors. Because this process originates from the $NV^-$ excited state — whose population is spin dependent — the resulting $NV^0$ emission inherits the same magnetic-field dependence observed for $NV^-$. In contrast, under 532 nm excitation, $NV^0$ is generated through processes that do not originate from the $NV^-$ excited state, leading to fluorescence that is largely insensitive to spin polarization.

These findings provide a framework for understanding charge-state conversion in NV ensembles in the presence of substitutional nitrogen. In particular, they demonstrate that $NV^0$ emission cannot be treated as a simple background, but rather depends on the microscopic mechanism by which it is produced. When generated through tunneling from $NV^-$, $NV^0$ becomes an integral part of the spin-dependent signal rather than a source of noise. This insight has practical implications for NV-based sensing: By choosing excitation conditions that preserve the spin-selective origin of $NV^0$, it becomes possible to utilize the full fluorescence spectrum for detection, thereby improving signal-to-noise and overall sensitivity. More broadly, the results highlight the importance of controlling both excitation wavelength and defect environment when optimizing ensemble-based sensors.

Looking ahead, the regime of shallow NV centers — central to most sensing applications — appears particularly intriguing[48]. In this case, the conversion efficiency from nitrogen to NV centers upon sample annealing is typically low, so that a large fraction of substitutional nitrogen remains in the lattice. As a result, many NV centers are expected to be in close proximity to one or more nitrogen donors, placing them naturally in the tunneling-dominated regime identified here[32]. This suggests that the interplay between $NV^-$, $NV^0$, and nearby donors may be even more pronounced for shallow ensembles than in bulk material. At the same time, shallow NVs are also subject to additional complexities, including the presence of surface-related defects and other charge traps. Understanding how these competing channels — donor-mediated tunneling, photoionization, and trapping — combine to determine the charge-state dynamics under illumination of varying wavelength remains an open question. Addressing this will be essential for developing a complete picture of near-surface NV behavior and for optimizing diamond materials for next-generation sensing applications.

### ACKNOWLEDGEMENTS

The authors thank Marco Capelli and Phillip Reineck for sharing their lifetime measurements, and Bryce Henson for his technical assistance, as well as Matt Sellars and Rose Ahlefeldt for helpful discussions. N.B.M. thanks the Australian Research Council for grant DP 170102232. M.W.D. is indebted to the Australian Research Council for award DE 170100169. C.A.M. acknowledges support from the National Science Foundation through award NSF-CMI-2506082. RU acknowledges funding by the Max-Planck Society.

### REFERENCES

[1] M.W. Doherty, N.B. Manson, P. Delaney, F. Jelezko, J. Wrachtrup, L.C.L. Hollenberg. "The nitrogen-vacancy colour centre in diamond", *Phys. Rep.* **528**, 1 (2013).

[2] N.B. Manson, J.P. Harrison, M.J. Sellars, "Nitrogen-vacancy center in diamond: Model of the electronic structure and associated dynamics", *Phys. Rev. B* **74**, 104303 (2006).

[3] C.L. Degen, F. Reinhard, P. Cappellaro, "Quantum sensing", *Rev. Mod. Phys.* **89**, 035002 (2017).

[4] G. Balasubramanian, I.Y. Chan, R. Kolesov, M. Al-Hmoud, J. Tisler, C. Shin, C. Kim, A. Wojeik, P.R. Hemmer, A. Krueger, T. Hanke, A. Leitenstorfer, R. Bratschitsch, F. Jelezko, J. Wrachtrup, "Nanoscale imaging magnetometry with diamond spins under ambient conditions", *Nature* **455**, 648 (2008).

[5] J.R. Maze, P.L. Stanwix, J.S. Hodges, S. Hong, J.M. Taylor, P. Cappellaro, L. Jiang, M.V. Gurudev Dutt, E. Togan, A.S. Zibrov, A. Yacoby, R.L. Walsworth, M.D. Lukin, "Nanoscale magnetic sensing with an individual electronic spin in diamond", *Nature* **455**, 644 (2008).

[6] T. Staudacher, F. Shi, S. Pezzagna, J. Meijer, J. Du, C.A. Meriles, F. Reinhard, J. Wrachtrup, "Nuclear magnetic resonance spectroscopy on a (5nm)3 volume of liquid and solid samples", *Science* **339**, 561 (2013).

[7] F. Dolde, H. Fedder, M.W. Doherty, T. Nöbauer, F. Rempp, G. Balasubramanian, T. Wolf, F. Reinhard, L.C.L. Hollenberg, F. Jelezko, J. Wrachtrup, "Electric-field sensing using single diamond spins", *Nat. Phys.* **7**, 459 (2011).

[8] D.J. McCloskey, N. Dontschuk, A. Stacey, C. Pattinson, A. Nadarajah, L.T. Hall, L.C.L. Hollenberg, S. Prawer, D.A. Simpson, "A diamond voltage imaging microscope", *Nat. Phot.* **16**, 730 (2022).

[9] D.M. Toyli, C.F. de las Casas, D.J. Christle, D.D. Awschalom, "Fluorescence thermometry enhanced by the quantum coherence of single spins in diamond", *Proc. Natl. Acad. Sci. U.S.A.* **110**, 8417 (2013).

[10] Y.Y. Hui, O.Y. Chen, T. Azuma, B.-M. Chang, F.-J. Hsieh, H.-C. Chang, "All-optical thermometry with nitrogen-vacancy centers in nanodiamond-embedded polymer films", *J. Phys. Chem. C* **123**, 15366 (2019).

[11] A. Laraoui, H. Aycock-Rizzo, X. Lu, Y. Gao, E. Riedo, C.A. Meriles, "Imaging thermal conductivity with nanoscale resolution using a scanning spin probe", *Nat. Commun.* **6**, 8954 (2015).

[12] M.W. Doherty, V.V. Struzhkin, D.A. Simpson, L.P. McGuinness, Y. Meng, A. Stacey, T.J. Karle, R.J. Hemley, N.B. Manson, L.C.L. Hollenberg, S. Prawer, "Electronic properties and metrology applications of the diamond $NV^-$ center under pressure", *Phys. Rev. Lett.* **112**, 047601 (2014).

[13] E. Bauch, S. Singh, J. Lee, C.A. Hart, J.M. Schloss, M.J. Turner, J.F. Barry, L.M. Pham, N. Bar-Gill, "Decoherence of ensembles of nitrogen-vacancy centers in diamond", *Phys. Rev. B* **102**, 134210 (2020).

[14] J. Achard, V. Jacques, A. Tallaire, "Chemical vapour deposition diamond single crystals with nitrogen-vacancy centres: a review of material synthesis and technology for quantum sensing applications", *J. Phys. D: Appl. Phys.* **53**, 313001 (2020).

[15] A.M. Edmonds, C.A. Hart, M.J. Turner, P.-O. Colard, J.M. Schloss, K. Olsson, R. Trubko, M.L. Markham, A. Rathmill, B. Horne-Smith, W. Lew, A. Manickam, S. Bruce, P.G. Kaup, J.C. Russo, M.J. DiMario, J.T. South, J.T. Hansen, D.J. Twitchen, R.L. Walsworth, "Generation of nitrogen-vacancy ensembles in diamond for quantum sensors: Optimization and scalability of CVD processes", arXiv 2004.01746 (2020).

[16] L. Razinkovas, M. Maciaszek, F. Reinhard, M.W. Doherty, A. Alkauskas, "Photoionization of negatively charged NV centers in diamond: Theory and ab initio calculations", *Phys. Rev. B* **104**, 235301 (2021).

[17] M.N.R. Ashfold, J.P. Goss, B.L. Green, P.W. May, M.E. Newton, C.V. Peaker, "Nitrogen in diamond", *Chem. Rev.* **120**, 5745 (2020).

[18] M. Capelli, L. Lindner, T. Luo, J. Jeske, H. Abe, S. Onoda, T. Ohshima, B. Johnson, D.A. Simpson, A. Stacey, "Proximal nitrogen reduces the fluorescence quantum yield of nitrogen-vacancy centres in diamond", *New J. Phys.* **24**, 033053 (2022).

[19] M. Mahdia, A. Lozovoi, J. Rovny, Z. Yuan, C.A. Meriles, N.P. de Leon, "High-efficiency, high-fidelity charge initialization of shallow nitrogen vacancy centers in diamond", *Phys. Rev. Appl.* **25**, 014006 (2026).

[20] A. Alkauskas, C.G. Van de Walle, L. Razinkovas, R. Ulbricht, "Charge state equilibration of nitrogen-vacancy center ensembles in diamond: The role of electron tunneling", arXiv 2512.00952 (2025).

[21] N. Aslam, G. Waldherr, P. Neumann, F. Jelezko, J. Wrachtrup, "Photo-induced ionization dynamics of the nitrogen-vacancy defect in diamond investigated by single-shot charge state detection", *New J. Phys.* **15**, 013064 (2013).

[22] C.X. Li, Q.Y. Zhang, N. Zhou, B.C. Hu, C.Y. Ma, C. Zhang, Z. Yi, "UV-induced charge-state conversion from the negatively to neutrally charged nitrogen-vacancy centers in diamond", *J. Appl. Phys.* **132**, 215102 (2022).

[23] A. Lozovoi, H. Jayakumar, D. Daw, G. Vizkelethy, E. Bielejec, J. Flick, M.W. Doherty, C.A. Meriles, "Optical activation and detection of charge transport between individual colour centres in diamond", *Nat. Electron.* **4**, 717 (2021).

[24] E. Bourgeois, M. Gulka, M. Nesládek, "Photoelectric detection and quantum readout of nitrogen-vacancy center spin states in diamond", *Adv. Optical Mater.* **8**, 1902132 (2020).

[25] A. Lozovoi, G. Vizkelethy, E. Bielejec, C.A. Meriles, "Imaging dark charge emitters in diamond via carrier-to-photon conversion", *Sci. Adv.* **8**, eabl9402 (2022).

[26] E. Bourgeois, A. Jarmola, P. Siyushev, M. Gulka, J. Hrubý, F. Jelezko, D. Budker, M. Nesládek, "Photoelectric detection of electron spin resonance of nitrogen-vacancy centres in diamond", *Nat. Commun.* **6**, 8577 (2015).

[27] H. Jayakumar, J. Henshaw, S. Dhomkar, D. Pagliero, A. Laraoui, N.B. Manson, R. Albu, M.W. Doherty, C.A. Meriles, "Optical patterning of trapped charge in nitrogen-doped diamond", *Nat. Commun.* **7**, 12660 (2016).

[28] N.B. Manson, M. Hedges, M.S.J. Barson, R. Ahlefeldt, M.W. Doherty, H. Abe, T. Ohshima, M.J. Sellars, "$NV^-$–$N^+$ pair centre in 1b diamond", *New J. Phys.* **20**, 113037 (2018).

[29] R. Giri, F. Gorrini, C. Dorigoni, C.E. Avalos, M. Cazzanelli, S. Tambalo, A. Bifone, "Coupled charge and spin dynamics in high-density ensembles of nitrogen-vacancy centers in diamond", *Phys. Rev. B* **98**, 045401 (2018).

[30] M. Capelli, L. Lindner, T. Luo, J. Jeske, H. Abe, S. Onoda, T. Ohshima, B. Johnson, D.A. Simpson, A. Stacey, "Proximal nitrogen reduces the fluorescence quantum yield of nitrogen-vacancy centres in diamond", *New J. Phys.* **24**, 033053 (2022).

[31] M.T. Luu, A.T. Younesi, R. Ulbricht, "Nitrogen-vacancy centers in diamond: discovery of additional electronic states", *Mater. Quantum Technol.* **4**, 035201 (2024).

[32] S. Dhomkar, H. Jayakumar, P.R. Zangara, C.A. Meriles, "Charge dynamics in near-surface, variable-density ensembles of nitrogen-vacancy centers in diamond", *Nano Lett.* **18**, 4046 (2018).

[33] A.T. Collins, "The Fermi level in diamond", *J. Phys.: Condens. Matter* **14**, 3743 (2002).

34 R. Monge, T. Delord, C.A. Meriles, "Reversible optical data storage below the diffraction limit", *Nat. Nanotech.* **19**, 202 (2024).

35 S. Dhomkar, J. Henshaw, H. Jayakumar, C.A. Meriles, "Long-term data storage in diamond", *Science Adv.* **2**, e1600911 (2016).

36 J.-P. Tetienne, L. Rondin, P. Spinicelli, M. Chipaux, T. Debuisschert, J.-F. Roch, V. Jacques, "Magnetic-field-dependent photodynamics of single NV defects in diamond: an application to qualitative all-optical magnetic imaging", *New J. Phys.* **14**, 103033 (2012).

37 N.D. Lai, D. Zheng, F. Jelezko, F. Treussart, J.-F. Roch, "Influence of a static magnetic field on the photoluminescence of an ensemble of nitrogen-vacancy color centers in a diamond single-crystal", *Appl. Phys. Lett.* **95**, 133101 (2009).

38 M.L. Goldman, M.W. Doherty, A. Sipahigil, N.Y. Yao, S.D. Bennett, N.B. Manson, A. Kubanek, M.D. Lukin, "State-selective intersystem crossing in nitrogen-vacancy centers", *Phys. Rev. B* **91**, 165201 (2015).

39 J.-P. Tetienne, H. Hingant, L. Rondin, A. Cavaillès, L. Mayer, G. Dantelle, T. Gacoin, J. Wrachtrup, J.-F. Roch, V. Jacques, "Magnetic-field-dependent photodynamics of single NV defects in diamond", *Phys. Rev. B* **87**, 235436 (2013).

40 M.L. Goldman, A. Sipahigil, M.W. Doherty, N.Y. Yao, S.D. Bennett, M. Markham, D.J. Twitchen, N.B. Manson, A. Kubanek, "Phonon-induced population dynamics and intersystem crossing in nitrogen-vacancy centers", *Phys. Rev. Lett.* **114**, 145502 (2015).

41 M. Nesládek, L.M. Stals, A. Stesmans, M. Vaněček, "Dominant defect levels in diamond thin films: A photocurrent and electron paramagnetic resonance study", *Appl. Phys. Lett.* **72**, 3306 (1998).

42 R.G. Farrer, "On the substitutional nitrogen donor in diamond", *Solid State Commun.* **7**, 685 (1969).

43 F.J. Heremans, G.D. Fuchs, C.F. Wang, R. Hanson, D.D. Awschalom, "Generation and transport of photo-excited electrons in single-crystal diamond", *Appl. Phys. Lett.* **94**, 152102 (2009).

44 J. Isberg, A. Tajani, D.J. Twitchen, "Photoionization measurement of deep defects in single-crystalline CVD diamond using the transient-current technique", *Phys. Rev. B* **73**, 245207 (2006).

45 K. Xu, D. Pagliero, G.I. Lopez Morales, J. Flick, A. Wolcott, and Carlos A. Meriles, "Photoinduced charge injection from shallow point defects in diamond into water", *ACS Appl. Mater. Interf.* **16**, 37226 (2024).

46 J. Rosa, M. Vaněček, M. Nesládek, L.M. Stals, "Photoionization cross-section of dominant defects in CVD diamond", *Diamond Relat. Mater.* **8**, 721 (1999).

47 R. Ulbricht, S.T. Van Der Post, J.P. Goss, P.R. Briddon, R. Jones, R.U.A. Khan, M. Bonn, "Single substitutional nitrogen defects revealed as electron acceptor states in diamond using ultrafast spectroscopy", *Phys. Rev. B* **84**, 165202 (2011).

48 R. Giri, C. Dorigoni, S. Tambalo, F. Gorrini, A. Bifone, "Selective measurement of charge dynamics in an ensemble of nitrogen-vacancy centers in nanodiamond and bulk diamond", *Phys. Rev. B* **99**, 155426 (2019).